# "Analysis of Seawater Quality Parameters and Treatment with Hydrodynamic Cavitation Method"

Divya Patil[1], Dr. Pankaj P. Gohil[2], *Dr. Hemangi Desai [1]

1. Shree Ramkrishna Institute of Computer Education and Applied Sciences, Sarvajanik University, Surat

2. Sarvajanik College of Engineering & Technology, Sarvajanik University, Surat

*hemangi.desai@srki.ac.in

**ABSTRACT**

*The aim of the research was to compare the quality parameters of Seawater before and after hydrodynamic cavitation treatment. The Hydrodynamic Cavitation Method for water treatment gives the highest reduction in Turbidity (100%), the second highest reduction in TSS (83.86%), and the lowest reduction in $Na+$ (8.47%), according to the study and analysis of various quality parameters.*

*When compared to CPCB Water Quality Criteria, treated water is suitable for outdoor bathing; it again satisfies the standards for classes SW-I, SW-II, SW-III, SW-IV and SW-V, i.e., treated sea water can be used for a variety of purposes, including bathing, contact water sports, commercial fishing, mariculture, ecologically sensitive zones, aesthetics, harbour, waters Navigation and Controlled water disposal.*

*The SAR value of treated water, which is 1.72, indicates it is appropriate for all types of soil and crops. The treated water's WQI, which is 65, showed that it is of Fair quality and may be used for industrial and irrigation uses. Water quality can be improved by recycling the treated water for an additional 24 hours using the hydrodynamic cavitation method. Recycling of one time treated sea water will result in higher-quality water that can be used for a variety of purposes.*

*The Hydrodynamic Cavitation method by using venturi - orifice is proven to be the most effective over the other methods. Because it does not require any chemical reagent, hence does not produce any hazardous chemical waste, and maintains an eco-friendly and economically sustainable, environment benign technique for the treatment of Seawater.*

**Key Words:** Hydrodynamic Cavitation Method, venturi - orifice, Water Quality Criteria, SAR value, WQI.

# 1. Introduction:

# 1 Seawater

More than 70% of the surface of the Earth is covered by seawater, the liquid that makes up the oceans and seas. 96.5 percent of seawater is made up of pure water, 2.5 percent salt, and minor amounts of dissolved inorganic and organic compounds, particles, and a few atmospheric gases. A wealth of chemical elements with industrial importance can be found in seawater. Sodium chloride, or table salt, is still produced by evaporating seawater in several places of the world. Additionally, desalinated seawater can provide an endless supply of drinkable water.[1] The average salinity of the oceans around the world is 3.5% (35 g/L, 53 ppt).This indicates that there are roughly 35 grammes of dissolved salts in every kilogramme, or roughly one litre by volume (mostly sodium (Na+) and chloride (Cl-) ions). At the surface, the density is 1.025 Kg/L on average. Due to the dissolved salts, which increase the mass by a bigger proportion than a volume, seawater has a higher density than both fresh water and clean water (1.0 Kg/L at 39 °F). As the concentration of salt rises, seawater's freezing point falls. It freezes at about -2° C (28° F) at average salinity.[2] In 2010, scientists discovered the 2.6°C (27.3° F) lowest seawater ever seen in a liquid condition in a stream beneath an Antarctic glacier. The pH range of seawater is normally restricted to 7.5 to 8.4.[3]

## 1.1 Properties

### 1.1.1 Salinity

Even though the most majority of seawater has a salinity of between 31 and 38 g/kg, or 3.1 and 3.8%, not all seawater is equally salty. The red sea is the saltiest open sea because of its extraordinarily high rates of evaporation, little precipitation, little river run-off, and restricted circulation. The absolute salinity of seawater has historically been approximated using a number of salinity scales. The "Practical Salinity Units (PSU)" scale was widely used. The current salinity standard is "Reference Salinity"[4], which expresses salinity in units of g/kg.

### 1.1.2 Density

Depending on the temperature and salinity, the density of surface saltwater varies from approximately 1020 to 1029 Kg/m3. Seawater has a density of 1023.6 kg/m3 at a temperature of 25 °C, a salinity of 35 g/kg, and a pressure of 1 atm.[5][6] Seawater can have a density of at least 1050 kg/m3 when it is deep underwater and subject to intense pressure. With salinity, saltwater density also varies. A typical seawater brine with a salinity of 120 g/kg at 25 °C and one atmosphere has a density of 1088 kg/m3.[7]

### 1.1.3 pH Value

Prior to the beginning of the Industrial Revolution (before 1850), the pH of the oceans' surface was around 8.2.Ocean acidification, a human-caused phenomenon related to carbon dioxide emissions, has been causing it to drop ever since. Between 1950 and 2020, the average pH of the ocean surface decreased about from 8.15 to 8.05.[8] The chemical makeup of saltwater makes it difficult to measure a pH, therefore there are several different pH scales used in chemical oceanography.[9] The pH range of seawater is normally restricted to 7.5 to 8.4.[10] There is no acknowledged reference pH-scale for seawater, and readings based on multiple scales can range by as much as 0.14 units.[11]

## 1.2 Chemical Composition

Compared to all other types of freshwater, seawater has around 2000 times more dissolved ions. Despite having around 2.8 times as much bicarbonate as river water, seawater has a much lower amount of bicarbonate as a proportion of total dissolved ions. The majority of the dissolved ions that are most prevalent in seawater are calcium, magnesium, sodium, and chloride.[12] About 70% of the earth's surface is made up of oceans and seas. They are around 15 times larger in volume than land. Along with traces of all types of universe-possible matter, seawater also contains 0.05M MgSO4 and 0.5M NaCl. According to an estimate, there is enough salt in the entire ocean to cover the continents with a layer 100 meters deep.

The elemental composition of seawater is shown in Table 1.[12]

**Table 1 : The Elemental composition of seawater**

| Element | Chemical species | Abundance mg/L (ppm) | Resident time (year) |
|---|---|---|---|
| Na | $Na^+$ | $10.5 \times 10^3$ | $6 \times 10^3$ |
| K | $K^+$ | 380 | $1.1 \times 10^7$ |
| Mg | $Mg^{2+}$, $MgSO_4$ | $1.35 \times 10^3$ | $4.5 \times 10^7$ |
| Ca | $Ca^{2+}$, $CaSO_4$ | 400 | $8.0 \times 10^6$ |
| Sr | $Sr^{2+}$, $SrSO_4$ | 8.0 | $1.1 \times 10^7$ |
| Ba | $Ba^{2+}$, $BaSO_4$ | 0.03 | $8.4 \times 10^4$ |
| Fe | $Fe(OH)_3$ | 0.01 | $1.4 \times 10^2$ |
| Zn | $Zn^{2+}$, $ZnSO_4$ | 0.01 | $1.8 \times 10^5$ |
| Cu | $Cu^{2+}$, $CuSO_4$ | 0.02 | $5.0 \times 10^4$ |
| Ni | $Ni^{2+}$, $NiSO_4$ | 0.02 | $1.8 \times 10^4$ |
| Mn | $Mn^{2+}$, $MnSO_4$ | 0.02 | $1.4 \times 10^4$ |
| V | $VO_2(OH)$ | 0.02 | $1.0 \times 10^4$ |
| Co | $Co^{2+}$, $CoSO_4$ | $5 \times 10^{-4}$ | $1.8 \times 10^4$ |
| Cd | $Cd^{2+}$, $CdSO_4$ | $11 \times 10^{-4}$ | $5.8 \times 10^5$ |

$Na^+$, $Cl^-$, and $Mg^{+2}$, the most prevalent elements (aside from O and H), make up 90% of the matter in saltwater, while $K^+$, $Ca^{+2}$, and $SO_4^{2-}$ are present in 3% and other elements make up 7%. The final sinks for many compounds involved in multiple geochemical processes are the seas and oceans. $Pb^{+2}$, $PbSO_4$, $Cs^{+2}$, $V^{+3}$, $AgCl_2$, $Cl^-$, $SO_4^{-2}$, $HCO^{-3}$, $H_2CO_3$, $B(OH)_3$, $HAsO_4^{-2}$, $H_3AsO_3$, and other chemical species found in saltwater also contribute.

## 1.3 Physical Chemistry of Seawater

Since it involves a system with unusual properties like an average temperature of 5°C **(0–30°C) and a pressure of 200 atm (1 atm at the surface and 1000 atm at the bottom),** seawater chemistry is quite challenging to understand. According to estimates, the oceans have rotated roughly 0.5 million times during the course of 500 million years, assuming a 1000-year rotation period. This demonstrated that the liquid phase had been fully included. The ongoing interactions of seawater with the atmosphere, biosphere, and sediment provide equilibrium processes a new dimensions.

### 1.4 Seawater Model

For seawater model, consider all constituents which have contributed to the formation of seawater. Goldschmidt estimated that for each kilogram of seawater, 600 g igneous rocks had been decomposed. Also one litre of seawater is in equilibrium with 0.6 kg of sediment and three liters of air.

### 1.4.1 Construction and Comparison of Seawater model with Real system

Seawater model (Fig.1) may be constructed by observing a process in which 1 L of seawater (with the corresponding amounts of solids and gas phase) is prepared by mixing with the constituents in order of their abundance till equilibrium is reached. This model is then compared with the real system that is, average seawater, sediments and air. Start with 1 L or 54.9 mole of water. Add elements like hydroxides or oxides. Shake vigorously till equilibrium is reached. Analyse solid and solution phase. It has been found that resemblances between the equilibrium model and real system is better than the expectation. Such a model gives useful information about the solid phase and chemical species in solution in sea / ocean water. [22]

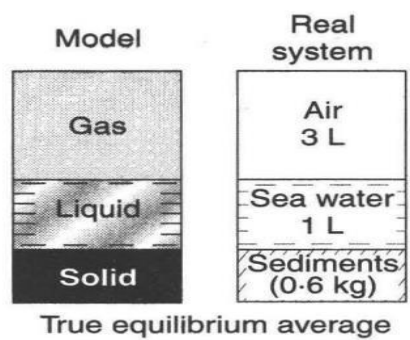

Fig 1: Seawater model compared with Real system

### 2. Seawater Analysis And Treatment

Seawater analysis and treatment is the process of analyzing the chemical composition of seawater and applying military science in place to make it appropriate for a variety of uses. For many applications, including industrial processes, marine research, environmental monitoring, and desalination for the production of freshwater, it is crucial to analyse and treat seawater. To understand the quality, content, and potential effects on the environment and human activities of seawater, it is necessary to investigate its physical, chemical, and biological property. Salinity, temperature, pH, dissolved oxygen, nutrients, heavy metals, organic compounds, and microbial parameters can be examined. To evaluate the quality of seawater, analytical methods like spectrophotometry, chromatography, titration, and microbiological assays are frequently applied. The first step in treating seawater is called desalination, which is the process of purifying freshwater by eliminating salt and other contaminants. Reverse osmosis is the most widely used desalination technique. In this process, seawater is pushed through a semipermeable membrane under pressure to remove salt and other impurities from water. Ion exchange, electrodialysis, and distillation are further desalination techniques and many more.[13-20] This paper provides an oversight of conventional and developing desalination technologies, emphasizing their existing state and subsequent potential to reduce water scarcity. Conventional hybrid desalination systems (NF-RO-MSF, MED-AD, FO-MED, MSF-MED, RO-MED, RO-MSF and RO-MD) are briefly discussed. This study reveals that the integration of solar thermal energy with desalination has a great potential to substantially reduce greenhouse emissions besides providing the quality and/or quantity of potable water in cost-effective ways. [21]

Depending on the intended use of the water, seawater treatment may also involve eliminating or removing particular pollutants. The analysis and treatment of seawater is essential for maintaining freshwater supplies, supporting a variety of industries, safeguarding marine life, and promoting public health.

### 3: Materials And Methods

#### 3:1 Sampling site
Dumas Beach, Surat

#### 3:2 Sample Collection
Sample was collected directly into two 20L Carboy container that have been rinsed three times with sample water from 20 – 30 cm below surface of water.

#### 3:3 Glass-wares
Pipettes (25ml, 10ml, 5ml, 1ml), Measuring Cylinder (100ml, 50ml, 10ml), Flasks (100ml, 50ml), Beakers (500ml, 250ml, 100ml), Test tubes, Standard Measuring Flask (SMF) (25ml), Evaporating Dish, BOD Bottles, Burette (50ml, 25ml), Glass rod Thermometer.

#### 3:4 Others
Bunsen Burner, Distilled Water, Whatman filter paper no.42, Test-tube stand, Tissue paper, Burette Stand.

#### 3:5 Instruments
Drying Oven, Dual Beam Spectrophotometer (Model No. EQ-824), Flame photometer (Model No. EQ 855A), Desiccator Aerator pH meter (Model No. EQ-615) Conductivity meter (Model No. 664A), COD Digester (Model No. EQ-831), COD Analyzer (Model No. EQ-831A)

#### 3:6 Methods to Determine Physicochemical parameters of Seawater
The analysis methods were used to determine the Seawater quality is shown in Table 2.

Table 2: Analysis method for Seawater quality parameters

| Parameter | Method | Reference[23] |
|---|---|---|
| pH Electrical Conductivity | Electrometric | APHA |
| TS, TSS, TDS | Oven Drying & Gravimetric | APHA |
| DO | Iodometric Titration | APHA |

| BOD | Winkler's Method | APHA |
|---|---|---|
| COD | Closed Reflux - Spectrophotometric | APHA |
| Cl$^-$ | Argentometric Titration | APHA |
| Total Hardness<br>Ca$^{+2}$ & Mg$^{+2}$ Hardness<br>Zn$^{+2}$, Cu$^{+2}$, Fe$^{+2}$ | Complexometric Titration | APHA |
| Total Acidity<br>Total Alkalinity<br>Salinity | Volumetric Titration | APHA |
| Turbidity | Spectrophototric | APHA |
| Na$^+$, K$^+$ | Flame Photometric | APHA |

### 3:7 Hydrodynamic Cavitation Method

Hydrodynamic Cavitation as a green and effective means of water treatment has attracted much attention. During the hydrodynamic cavitation, enormous energy could be released into the surrounding liquid which caused thermal effect, mechanical effects and chemical effects (formation of (OH$^o$) - Hydroxyl free radicals). These conditions can degrade bacteria and organic substance in sewage. Moreover the combination of hydrodynamic cavitation and other water treatment methods can produce coupling effect.[24] Hydrodynamic cavitation is a promising application in wastewater treatment due to its simple reactor design and capacity in large scale operation. Easy large-scale operation, effective combination with intensified strategies and capability to deal with toxic compound contribute to the great potential of hydrodynamic cavitation.[25]

### 3:8 Seawater Treatment by using Cavitation Method

As cavitation method was selected for Seawater treatment collected from Dumas Beach, Sura.t It is needed to setup a small-scale hydrodynamic cavitation reactor for which a cavitation producing devices has been selected. The Venturi tube and an Orifice plate has been selected as cavitation producing devices for hydrodynamic cavitation reactor. Cavitation is a phenomenon in which the static pressure of liquid reduces to below the liquid's vapour pressure, leading to the formation of small-vapour filled cavities in the liquid. When subjected to higher pressure, these cavities called, "bubbles" or "voids", collapse and can generate shock waves that may damage machinery.[26] Hydrodynamic Cavitation is one of the advanced oxidation process in which (OH$^o$) - Hydroxyl free radicals and Superoxide radicals are used but majorly (OH$^o$) radicals are used. (OH$^o$) radicals are strong oxidant, so they oxidize the organic matter present in the water.[27] Hydrodynamic Cavitation involves three mechanisms: nucleation, bubble growth and bubble implosion.[28] Hydrodynamic Cavitation is the process of vaporisation, bubble generation and bubble implosion which occurs in flowing liquid as a result of a decrease and subsequent increase in local pressure.[26] When the cavitation bubbles collapse, a large amount of energy is released, which affects the surrounding liquid environment in the form of thermal effect, mechanical effect and chemical effect.[24] The liquid environment generated by hydrodynamic cavitation can be used effectively for water treatment. When the liquid flows through the throat of the venturi tube, the decrease of cross-sectional area leads to the increase of the flow velocity, which reduces the local pressure of liquid.[29] When the local pressure falls below the corresponding saturated vapour pressure, cavitation bubbles are generated and grow. As liquid continues to flow, the pressure returns to normal, cavitation bubbles collapse and cavitation phenomenon occurs. When the liquid flows through the orifice of orifice plate of hydrodynamic cavitation reactor, the sudden decrease of cross sectional area leads to increase of liquid pressure, resulting in hydrodynamic cavitation phenomenon.[30] The orifice place in hydrodynamic cavitation reactor is used commonly because of its simple structure, convenient fabrication, low price and high stability.

### 3:8:1 Experimental Setup

Following is the experimental setup for hydrodynamic cavitation reactor for Seawater treatment by using cavitation method (Fig. 2,3,4). The setup consists of an orifice plate, venturi tube that operates at atmospheric pressure to be used. The tank (20 L) is used for water storage and collection. The pump is used for circulation of Seawater and valves are used for regulating the flow.

- System is compatible for pressure range up to 3 bar pressure and temperature up to 60° C
- Investigation of cavitation processes in a Venturi nozzle of ¾" connection of PP material
- Venturi nozzle with 3 pressure measuring points
- Orifice plates of 2 mm, 4 mm, 6 mm and 8 mm hole size with 2 pressure measuring points
- Adjustment of the flow rate via hall valves
- Temperature controller for controlling the temperature of tank water
- Flow measurement using Rotameter of 250 – 2500 L/h measuring range

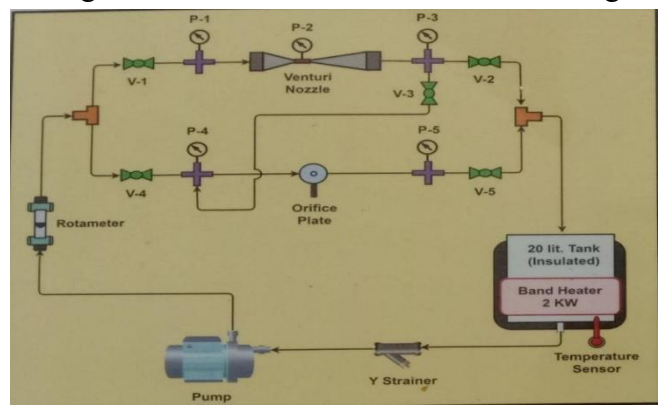

**Fig 2: Setup of Hydrodynamic Cavitation Reactor**

**Cavitation refers** to the formation of vapour bubbles in flowing fluids due to strong low pressure. As the flow velocity increases, the static pressure of fluid falls to vapour pressure and leads to the formation of vapour bubbles in narrowest cross-connection. The bubbles are carried along by the flow and implode if, with decreasing velocity, the static pressure rises above the vapour pressure of the fluid.

**Venturi nozzle** is made of Poly-propylene material. There are three pressure measuring points on Venturi nozzle: at the inlet, at the narrowest point and at outlet. The flow rate and the pressures can be adjusted via two ball valves which are located at the inlet and outlet of the pipe system. The pressure distribution within the venturi nozzle is shown on three pressure gauges. The flow can be read off a rotameter.

**Orifice plates** are the most commonly used primary elements for differential pressure flow measurement. Orifice plates are inserted within a circular pipe, they create an obstacle, increase the speed of the fluid and generate a pressure difference between upstream and downstream of the restriction. This differential pressure measurement is proportional to the flow rate value.

After collecting the sample of Seawater, the water is treated by cavitation method by venturi nozzle and orifice plate for 24 hrs.

The reading for the same were recorded as shown in Table 3.

**Table 3: Readings of Hydrodynamic Cavitation Reactor**

| Readings | Temperature of water | Volume of Raw Seawater | Flow rate | P1 | P2 | P3 | P4 | Volume of Treated Seawater |
|---|---|---|---|---|---|---|---|---|
| Initial readings | 29.4°C (At 25°C Vapour pressure of water is 0.4578 Psi) | 15 L | 1000 lph | 0 Kg/Cm² (0 Psi) | -11.25 Kg/cm² (-360 Psi) | 0 Kg/cm² (0 Psi) | 0 Kg/cm² (0 Psi) | - |
| Final readings | 66.6°C (At 66°C Vapour pressure of water is 3.775 Psi) | 15 L | 1000 lph | 0 Kg/cm² (0 Psi) | -11.25 Kg/cm² (-275 Psi) | 0 Kg/cm² (0 Psi) | 0 Kg/cm² (0 Psi) | 5 L |

Physicochemical parameters of Seawater before and after treatment with Hydrodynamic Cavitation reactor were analyzed(Fig.5). Then the same parameters values were compared with the parameters values of untreated (raw) Seawater. Hence, the results of all parameters values of untreated Seawater were compared with parameters values of treated Seawater with cavitation method.

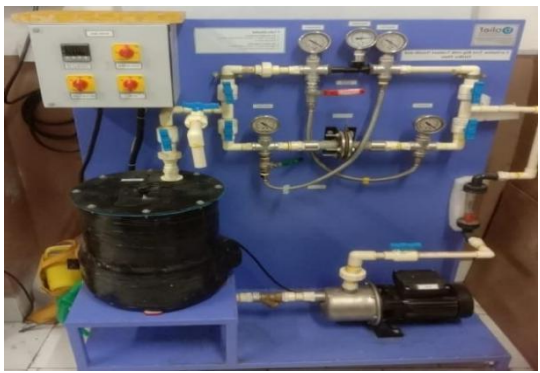
Fig 3: Physical setup of Hydrodynamic Cavitation reactor

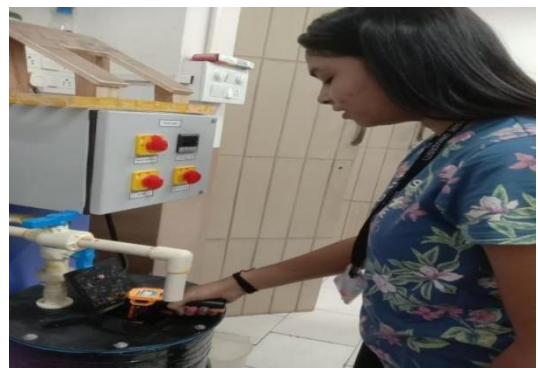
Fig 4: Checking of initial temperature of water before treatment

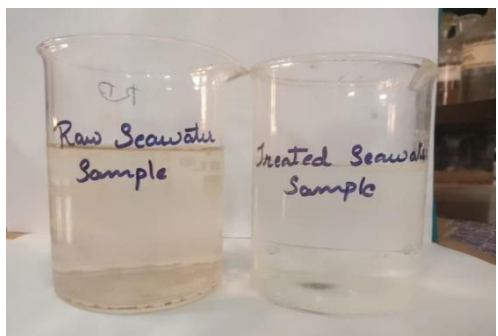
Fig 5: Raw and Treated Seawater sample

### 3:9 Formulas used for Calculation

The following parameters were taken into consideration to justify treated seawater quality . [31,32,33,34]

**1. % Reduction**

$$\% \text{ Reduction} = \frac{raw\ (mg/L) - treated\ (mg/L)}{raw\ (mg/L)} \times 100$$

**2. Sodium Absorption Ration (SAR )**

$$SAR = \frac{Na^+}{\sqrt{Ca^{2+} + Mg^{2+}/2}}$$

**3. Water Quality Index (WQI)**

Calculation of WQI was carried out by Horton's method.
The WQI is calculated using the following formula :
$$WQI = \sum q_n W_n / \sum W_n$$
Where, $q_n$ = Quality rating of $n^{th}$ water quality parameter.
$W_n$ = Unit weight of $n^{th}$ water quality parameter.

• **Quality rating ($q_n$)**

The quality rating ($q_n$) is calculated using the following equation :
$q_n = [(V_n - V_{id}) / (S_n - V_{id})] \times 100$
Where, $V_n$ = Estimated value of $n^{th}$ water quality parameter at a given sample location.
$V_{id}$ = Ideal value for $n^{th}$ parameter in pure water ($V_{id}$ pH = 7 and 0 for all other parameters)
$S_n$ = Standard permissible value for $n^{th}$ water quality parameter.

• **Unit weight**

The unit weight ($W_n$) is calculated by using the following equation :

$W_n = k / S_n$; Where, $S_n$ = Standard permissible value of $n^{th}$ water quality parameter

K = Constant of proportionality and it is calculated by using the following equation :

$k = [ 1 / ( \sum 1/S_{n=1,2,....n})]$

### 4: Result And Discussion

All Physicochemical parameters are analyzed before and after treatment of Seawater and they were compared with each other and their % Reduction is calculated.

**Table 4: Seawater Quality Analysis after and before treatment with Hydrodynamic Cavitation method**

| Seawater Quality Parameters | Raw (mg/L) | Treated (mg/L) | % Reduction |
|---|---|---|---|
| Turbidity | 6 NTU | 0 NTU | 100 |
| TSS | 4640 | 760 | 83.86 |
| BOD | 7.2 | 4 | 80 |
| Alkalinity | 50 | 10 | 80 |
| $Fe^{+2}$ | 390.95 | 111.7 | 71.42 |
| Acidity | 50 | 20 | 60 |
| DO | 12.2 | 5 | 59.01 |
| TS | 44,440 | 25,120 | 43.47 |
| COD | 342 | 198 | 42.1 |
| $Zn^{2+}$ | 13,076 | 7845.6 | 40 |
| $Cu^{2+}$ | 63.54 | 38.12 | 40 |
| TDS | 39,800 | 24,360 | 38.71 |
| $Ca^{2+}$ Hardness | 220 | 150 | 31.81 |
| Total Hardness | 520 | 400 | 23.07 |
| $Cl^-$ | 24,494.47 | 20,145.45 | 17.75 |
| Salinity | 44.250 | 36.393 | 17.75 |
| $Mg^{2+}$ Hardness | 300 | 250 | 16.66 |
| pH | 7.45 | 6.31 | 15.3 |
| Conductivity | 40.6 mS | 34.9 mS | 14.03 |
| $K^+$ | 800 | 700 | 12.5 |
| $Na^+$ | 17,700 | 16,200 | 8.47 |

### Table 5: Designated Best Use Water Quality Criteria[35]

| Designated-Best-Use | Class of water | Criteria |
|---|---|---|
| Drinking Water Source without conventional treatment but after disinfection | A | Total Coliforms Organism MPN/100ml shall be 50 or less; pH between 6.5 and 8.5; Dissolved Oxygen 6mg/l or more; Biochemical Oxygen Demand 5 days 20C 2mg/l or less |
| Outdoor bathing (Organised) | B | Total Coliforms Organism MPN/100ml shall be 500 or less pH between 6.5 and 8.5 Dissolved Oxygen 5mg/l or more; Biochemical Oxygen Demand 5 days 20C 3mg/l or less |
| Drinking water source after conventional treatment and disinfection | C | Total Coliforms Organism MPN/100ml shall be 5000 or less pH between 6 to 9 Dissolved Oxygen 4mg/l or more; Biochemical Oxygen Demand 5 days 20C 3mg/l or less |
| Propagation of Wild life and Fisheries | D | pH between 6.5 to 8.5 Dissolved Oxygen 4mg/l or more; Free Ammonia (as N) 1.2 mg/l or less |
| Irrigation, Industrial Cooling, Controlled Waste disposal | E | pH betwwn 6.0 to 8.5; Electrical Conductivity at 25C micro mhos/cm Max.2250; Sodium absorption Ratio Max. 26; Boron Max. 2mg/l |

### Table 6: Primary Water Quality Criteria For Bathing Water[35]

**PRIMARY WATER QUALITY CRITERIA FOR BATHING WATER**
(Water used for organised outdoor bathing)

| | CRITERIA | | RATIONALE |
|---|---|---|---|
| 1. Fecal Coliform MPN/100 ml: | 500 (desirable) 2500 (Maximum Permissible) | | To ensure low sewage contamination. Fecal coliform and fecal streptococci are considered as they reflect the bacterial pathogenicity. |
| 2. Fecal Streptococci MPN/100 ml: | 100 (desirable) 500 (Maximum Permissible) | | The desirable and permissible limits are suggested to allow for fluctuation in environmental conditions such as seasonal change, changes in flow conditions etc. |
| 2. pH: | Between 6.5 –8.5 | | The range provides protection to the skin and delicate organs like eyes, nose, ears etc. which are directly exposed during outdoor bathing. |
| 3. Dissolved Oxygen: | 5 mg/l or more | | The minimum dissolved oxygen concentration of 5 mg/l ensures reasonable freedom from oxygen consuming organic pollution immediately upstream which is necessary for preventing production of anaerobic gases (obnoxious gases) from sediment. |
| 4. Biochemical Oxygen demand 3 day,27°C: | 3 mg/l or less | | The Biochemical Oxygen Demand of 3 mg/l or less of the water ensures reasonable freedom from oxygen demanding pollutants and prevent production of obnoxious gases"; |

From the above tables 5-6, it is concluded that the Seawater after treatment can be used for Outdoor bathing water when compared to Primary Water Quality Criteria for Bathing Water by CPCB.

**Water Quality Standards for Coastal Waters Marine Outfalls (Table 7 to 12)**

In a coastal segment marine water is subjected to several types of uses. Depending on the types and uses, water quality criteria have been specified to determine it suitability for a particular purpose. Among the various types of uses there is one use that demands highest level of water quality / purity and that is termed a " Designated best use " in that stretch of the coastal segment. Based on this, primary water quality criteria have been specified for the following five designated best use.

### Table 7: Primary Water Quality Class and Designated Best Uses

| Class | Designated Best Use |
|---|---|
| SW – I | Salt pans, Shell fishing, Mariculture and Ecologically Sensitive Zone. |
| SW – II | Bathing, Contact Water Sports and Commercial fishing. |
| SW – III | Industrial cooling, Recreation (non-contact) and Aesthetics |
| SW – IV | Harbour |
| SW - V | Navigation and Controlled Waste Disposal |

### Table 8: Primary Water Quality Criteria for SW-I Waters

**Table 1.1 Primary Water Quality Criteria For Class SW-I Waters**
(For Salt pans, Shell fishing, Mariculture and Ecologically Sensitive Zone)

| S. No. | Parameter | Standards | Rationale/Remarks |
|---|---|---|---|
| 1. | pH range | 6.5-8.5 | General broad range, conducive for propagation of aquatic lives, is given. Value largely dependant upon soil-water interaction. |
| | | | (Contd.....) |
| 2. | Dissolved Oxygen | 5.0 mg/l or 60 percent saturation value, whichever is higher. | Not less than 3.5 mg/l at any time of the year for protection of aquatic lives. |
| 3. | Colour and Odour | No noticeable colour or offensive odour. | Specially caused by chemical compounds like creosols, phenols, naptha, pyridine, benzene, toluene etc. causing visible colouration of salt crystal and tainting of fish flesh. |
| 4. | Floating Matters | Nothing obnoxious or detrimental for use purpose. | Surfactants should not exceed an upper limit of 1.0 mg/l and the concentration not to cause any visible foam. |
| 5. | Suspended Solids | None from sewage or industrial waste origin | Settleable inert matters not in such concentration that would impair any usages specially assigned to this class. |
| 6. | Oil and Grease (including Petroleum Products) | 0.1 mg/l | Concentration should not exceed 0.1 mg/l as because it has effect on fish egs and larvae. |
| 7. | Heavy Metals: Mercury (as Hg) Lead (as Pb) Cadmium (as Cd) | 0.01 mg/l 0.01 mg/l 0.01 mg/l | Values depend on: (i) Concentration in salt, fish and shell fish. (ii) Average per capita consumption per day. (iii) Minimum ingestion rate that induces symptoms of resulting diseases. |

When treated Seawater's quality parameters are compared to the Primary Water Quality Criteria, all parameters—except heavy metals—meet the Quality Standards. Oil & Grease was not found in raw Seawater. Treated filtered Seawater was found transparent clear water. Hence treated Seawater fulfil the criteria of SW-I class and can be used for Salt pans, Shell fishing, Mariculture and Ecologically sensitive zone.

| Table 9: Primary Water Quality Criteria for Class SW-II Waters | Table 10: Primary Water Quality Criteria for Class SW-III Waters |
|---|---|

**Table 1.2 Primary Water Quality Criteria for Class SW-II Waters**
(For Bathing, Contact Water Sports and Commercial Fishing)

| S. No. | Parameter | Standards | Rationale/Remarks |
|---|---|---|---|
| 1. | pH range | 6.5-8.5 | Range does not cause skin or eye irritation and is also conducive for propagation of aquatic life. |
| 2. | Dissolved Oxygen | 4.0 mg/l or 50 percent saturation value whichever is higher. | Not less than 3.5 mg/l at anytime for protection of aquatic lives. |
| 3. | Colour and Odour | No noticeable colour or offensive odour. | Specially caused by chemical compounds like creosols phenols, naptha, benzene pyridine, volume etc. causing visible colouration of water and tainting of and odour in fish flesh. |
| 4. | Floating Matters | Nothing obnoxious or detrimental for use purpose. | None in concentration that would impair usages specially assigned to this class. |
| 5. | Turbidity | 30 NTU (Nephelo Turbidity Unit) | Measured at 0.9 depth. |
| 6. | Fecal Coliform | 100/100 ml (MPN) | The average value not exceeding 200/100 ml. in 20 percent of samples in the year and in 3 consecutive samples in monsoon months. |
| 7. | Biochemical Oxygen Demand (BOD) (3 days at 27°C) | 3 mg/l | Restricted for bathing (aesthetic quality of water). Also prescribed by IS:2296-1974. |

**Table 1.3 Primary Water Quality Criteria for Class SW-III Waters**
[For Industrial cooling, Recreation (non-contact) and Aesthetics]

| S. No. | Parameter | Standards | Rationale/Remarks |
|---|---|---|---|
| 1. | pH range | 6.5-8.5 | The range is conducive for propagation of aquatic species and restoring natural system. |
| 2. | Dissolved Oxygen | 3.0 mg/l or 40 percent saturation value whichever is higher. | To protect aquatic lives. |
| 3. | Colour and Odour | No noticeable colour or offensive odour. | None in such concentration that would impair usages specifically assigned to this class. |
| 4. | Floating Matters | No visible/obnoxious floating debris, oil slick, scum. | As in (3) above. |
| 5. | Fecal Coliform | 500/100 ml (MPN) | Not exceeding 1000/100 ml in 20 percent of samples in the year and in 3 consecutive samples in monsoon months. |
| 6. | Turbidity | 30 NTU | Reasonably clear water for Recreation, Aesthetic appreciation and Industrial cooling purposes. |
| *7. | Dissolved Iron (as Fe) | 0.5 mg/l or less | It is desirable to have the collective concentration of dissolved Fe and Mn less or equal to 0.5 mg/l to avoid scaling effect. |
| *8. | Dissolved Manganese (as Mn) | 0.5 mg/l or less | |

When Seawater quality parameters of treated water are compared with the Primary Water Quality Criteria of Class SW-II Waters, it completely fulfil the criteria of Standard Values of Class SW-II Waters. Hence, treated water can be used for Bathing, Contact Water Sports and Commercial Fishing. Treated seawater when compared with Primary Water Quality Criteria of Class SW-III Water, it fulfil the standard criteria for pH, Turbidity, DO and Colour and Odour. Hence, as per the Primary Water Quality Criteria of Class SW-III Water, treated water may be use for Industrial cooling, Recreation (non-contact) and Aesthetics.

| Table 11: Primary Water Quality Criteria of Class SW-IV Waters | Table 12: Primary Water Quality Criteria of Class SW-V Waters |
|---|---|

**Table 1.4 Primary Water Quality Criteria for Class SW-IV Waters**
(For Harbour Waters)

| S. No. | Parameter | Standards | Rationale/Remarks |
|---|---|---|---|
| 1. | pH range | 6.5-9.0 | To minimize corrosive and scaling effect. |
| 2. | Dissolved Oxygen | 3.0 mg/l or 40 percent saturation value whichever is higher | Considering bio-degradation of oil and inhibition to oxygen production through photosynthesis. |
| 3. | Colour and Odour | No visible-colour or offensive odour. | None from reactive chemicals which may corrode paints/metallic surfaces. |
| 4. | Floating materials Oil, grease and scum (including Petroleum products) | 10 mg/l | Floating matter should be free from excessive living organisms, which may clog or coat operative parts of marine vessels/equipment. |
| 5. | Fecal Coliform | 500/100 ml (PAN) | Not exceeding 1000/100 ml in 20 percent of samples in the year and in 3 consecutive samples in monsoon months. |
| 6. | Biochemical Oxygen Demand (3 days at 27°C) | 5 mg/l | To maintain water relatively free from pollution caused by sewage and other decomposable wastes. |

**Table 1.5 Primary Water Quality Criteria for Class SW-V Waters**
(For Navigation and Controlled Waste Disposal)

| S. No. | Parameter | Standards | Rationale/Remarks |
|---|---|---|---|
| 1. | pH range | 6.0-9.0 | As specified by New England Interstate Water Pollution Control Commission. |
| 2. | Dissolved Oxygen | 3.0 mg/l or 40 percent saturation value which ever is higher | To protect aquatic lives. |
| 3. | Colour and Odour | None is such concentration that would impair any usages specifically assigned to this class. | As in (1) above |
| 4. | Sludge deposits, Solid refuse floating oil, grease & scum. | None except for such small solids, amount that may result from discharge of appropriately treated sewage and/or individual waste effluents. | As in(1) above |
| 5. | Fecal Coliform | 500/100 ml (MPN) | Non exceeding 1000/100 ml in 20 percent of samples in the year and in 3 consecutive samples in monsoon months. |

(Contd.....)

104

(Contd.....)

Treated water fulfil the standard criteria for pH, DO, Colour and Odour and BOD of Class SW-IV Waters. Hence, treated water, as per the Primary Water Quality Criteria of Class SW-IV Waters can be used for Harbour Waters. Treated filtered Seawater was found transparent clear water. Treated water when compared with the Primary Water Quality Criteria of Class SW-V Waters, fulfil the standard criteria. Hence, treated water can be used for Navigation and Controlled waste disposal.

## 4:1 Analysis of Raw & Treated Seawater Quality Parameter

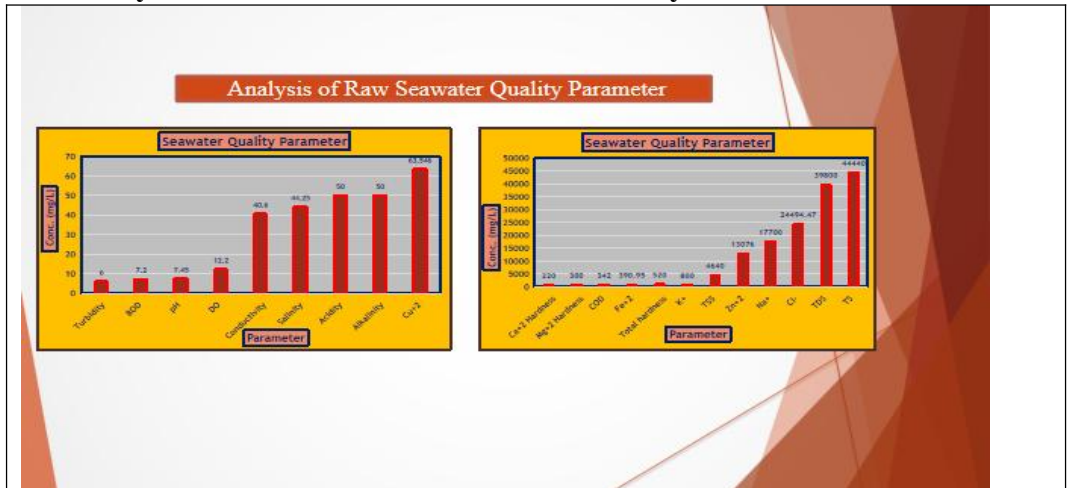

**Fig 6: Seawater Quality Parameters (Raw)**

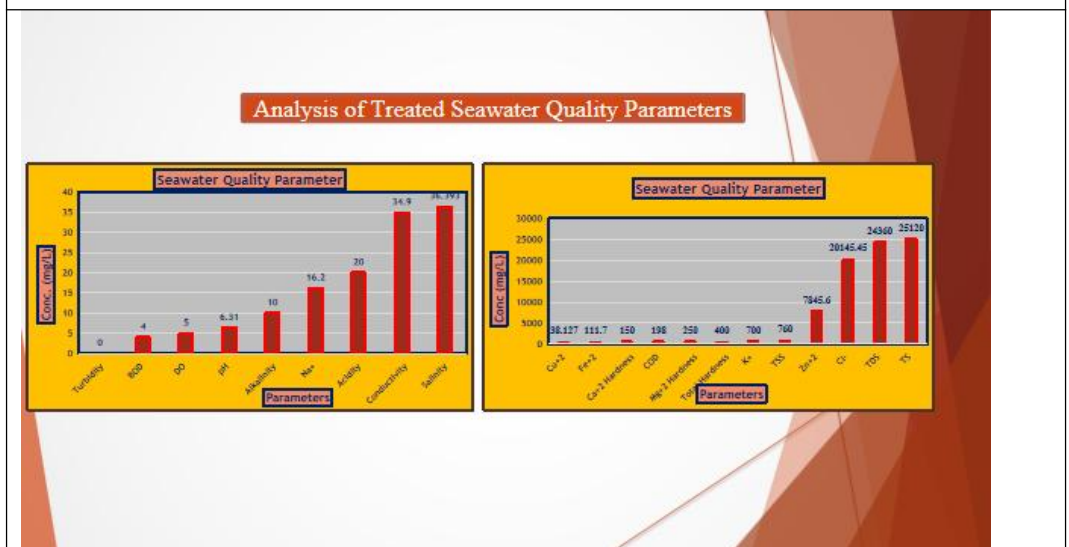

**Fig 7: Seawater Quality Parameter (Treated) by Hydrodynamic Cavitation Method for 24 Hrs**

## 4:2 Comparison of Raw and Treated Seawater Quality

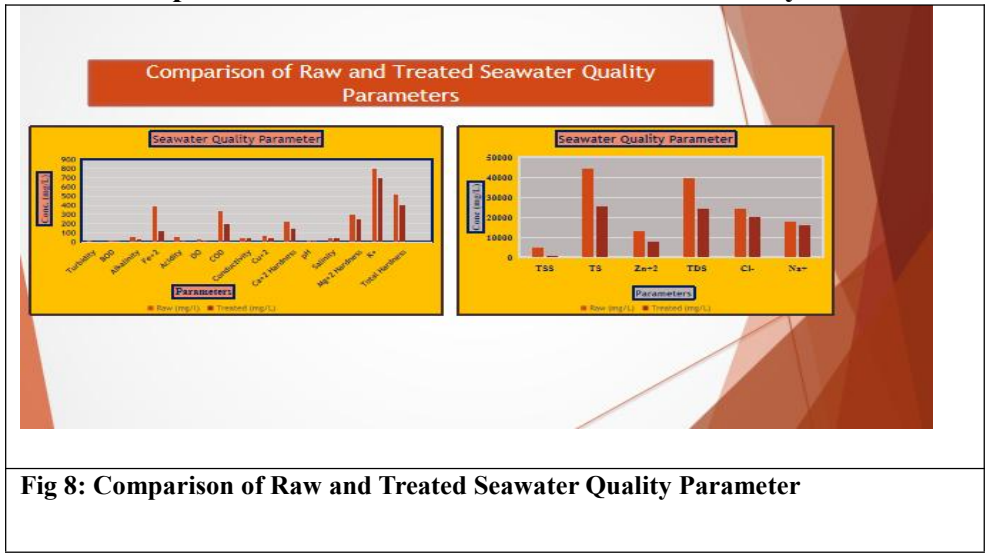

**Fig 8: Comparison of Raw and Treated Seawater Quality Parameter**

## 4:3 % Reduction of Seawater Quality Parameter after treatment by Hydrodynamic Cavitation Method for 24 Hrs.

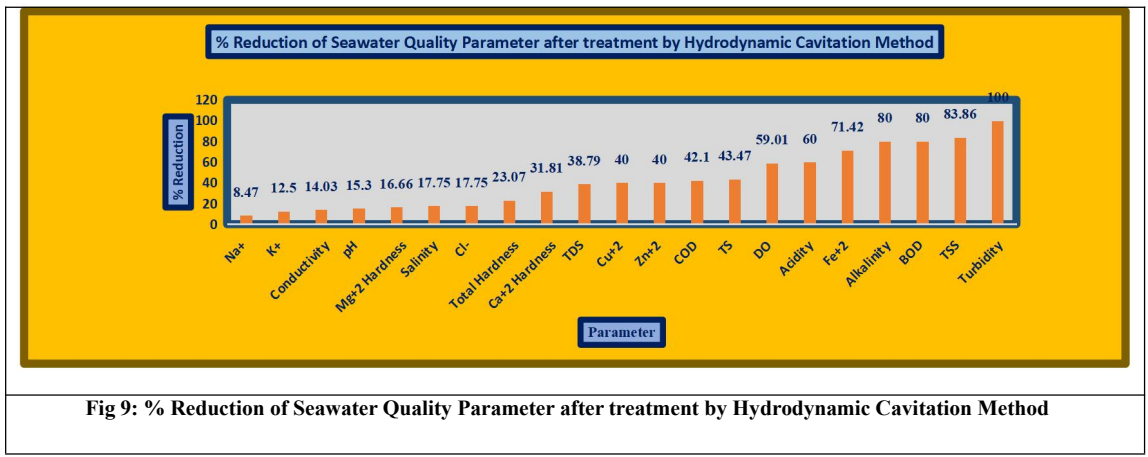

**Fig 9: % Reduction of Seawater Quality Parameter after treatment by Hydrodynamic Cavitation Method**

The result of analysis of raw Seawater Quality Parameters and treated Seawater Quality Parameters is shown in Table 4 and Fig 15 and 16. The comparison of raw and treated Seawater Quality Parameters is shown in Fig 17. % Reduction of Seawater Water Quality Parameters is shown in Fig 18.

From the Results of the Analysis done in this study, it is evident that the lowest reduction is obtained in the following parameters: Na+ (8.47%), $K^+$ (12.5%), electrical conductivity (14.03%), pH (15.3%), $Mg^{2+}$ Hardness (16.66%), Salinity (17.75%), $Cl^-$ (17.75%), Total Hardness (23.07), $Ca^{+2}$ (31.81), and TDS (38.79%).

Highest reduction is obtained in Turbidity (100%), TSS (83.86%), BOD (80%), Alkalinity (80%), $Fe^{3+}$ (71.42%), Acidity (60%) TS (43.47%), COD (42.1%), $Zn^{+2}$ (40%) and $Cu^{+2}$ (40%).

Due to the force of shock waves and decrease and subsequent increase in local pressure, there is a breakdown of colloidal aggregation i.e., Turbidity. Therefore after water treatment, highest reduction is achieved in Turbidity. Then after, the second highest reduction is obtained in TSS and about 40% reduction is gained for TDS. This may be depend upon the particle size in dissolved state. Upon certain particle size this force of shock waves and pressure is able to breakdown the particles. But for a very smaller size and micro size particles, it is not able to breakdown. Therefore, very less reduction is obtained for smaller size dissolved solids. During the hydrodynamic cavitation, enormous energy could be released into the surrounding liquid which caused thermal effect, mechanical effects and chemical effects. These conditions can degrade bacteria and organic substance in sewage.[24]

The Sodium Absorption Ration (SAR) value of treated Seawater obtained is 1.72.

$$SAR = \frac{8.47}{\sqrt{31.81+16.66/2}} = 1.72$$

**Table 13: Suitability of water with different values of SAR by CPCB**

| SAR | Suitability |
|---|---|
| 0 – 10 | For all types of soils and crops. |
| 10 – 18 | Coarse textured or organic soil with good permeability. Unsuitable for fine soil. |
| 18 – 26 | Harmful for all types of soils, requires good drainage, high leaching and gypsum addition. |
| > 26 | Unsuitable for irrigation purpose. |

From the above table 13, it is concluded that the treated water is suitable for all types of soil and crops.

Water Quality Index (WQI) is calculated by using the equation :

$$q_n = [(V_n - V_{id}) / (S_n - V_{id})] \times 100$$

and then compared with the Corresponding status of Water Quality by CPCB (Table 14) The Water Quality Index obtained for treated water is 65.

**Table 14: WQI and corresponding water quality status**

| No: | WQI | Status | Possible Usages |
|---|---|---|---|
| 1 | 0 – 25 | Excellent | Drinking, Irrigation and Industrial |
| 2 | 25 – 50 | Good | Drinking, Irrigation and Industrial |
| 3 | 51 – 75 | Fair | Irrigation and Industrial |

| 4 | 76 – 100 | Poor | Irrigation |
| 5 | 101 – 150 | Very Poor | Restricted use for Irrigation |
| 6 | Above 150 | Unfit for drinking | Proper treatment required before use. |

The water Quality Index (WQI) of treated water obtained is 65 i.e., Fair quality water is obtained after treatment by Hydrodynamic Cavitation Method. Hence, Treated Seawater can be used for Irrigation and Industrial usages. (Table 14) By using the hydrodynamic cavitation method, it is possible to recycle the treated water for another 24 hours to get improved water quality. Water recycling can be done repeatedly to provide better-quality water for a variety of uses.

## 5. Conclusion

By Hydrodynamic Cavitation Method, highest reduction is obtained in Turbidity (100%), second highest in TSS (83.86%) and lowest reduction in Na+ (8.47%). From SAR value i.e., 1.72 it is concluded that the treated seawater is suitable for all soils and crops. From WQI i.e., 65 it is concluded that the water obtained is of Fair Quality and is useful for Irrigation and Industrial purposes. It fulfils Primary Quality Criteria of Marine Outfalls and is useful for various purposes. The venturi-orifice cavitation approach has been shown to be more effective than the others, because it doesn't require any chemical reagent, creates no harmful chemical waste and treats seawater in a way that is both environmentally friendly and economically viable. Using Hydrodynamic Cavitation, it is possible to recycle the treated sweater for another 24 hours to get improved water quality. Water recycling can be done repeatedly to provide better quality water for various uses.